\documentclass[journal,onecolumn,draftclsnofoot]{IEEEtran}

\usepackage{graphicx}
\usepackage{multirow}
\usepackage{comment}
\usepackage{eurosym}
\usepackage{hhline}
\usepackage{amssymb}
\usepackage{amsmath}
\usepackage{setspace}
\usepackage{algorithm} 
\usepackage{algpseudocode}
\usepackage{textcomp}
\usepackage{subcaption} 
\usepackage{balance}
\usepackage{siunitx}
\usepackage{float}
\usepackage[table,xcdraw]{xcolor}
\usepackage{soul}
\renewcommand{\thetable}{\arabic{table}}

\ifCLASSINFOpdf
\else
\fi

\begin{document}

\title{Hypothesizing the Transmission of Smells: Six Methods of Chemical Travel}

\author{\IEEEauthorblockN{Usman Ahmad} \\
\IEEEauthorblockA{
1-School of Computer Science and Communication Engineering, Jiangsu University, China\\
2-Department of Computational Science, The University of Faisalabad, Pakistan \\
usman715@gmail.com
}
}

\markboth{}%
{Ahmad \MakeLowercase{\textit{et al.}}: Hypothesizing the Transmission of Smells: Six Methods of Chemical Travel"}

\maketitle

\begin{abstract}

In this paper, we propose a hypothesis regarding the travel and movement of chemicals between locations. We introduce six distinct methods to explain this process. The chemicals referred to in this article are those that we detect by the nose and can travel through the air. These are generally known as odorants or volatile compounds. Method 1 involves decomposing chemicals and transmitting their formulas digitally. Method 2 embeds chemicals in water, freezes it into ice, and transports it to the receiver, where the chemicals are extracted upon melting. Method 3 transfers chemicals via water logistics, with the receiver extracting them from the water. Method 4 uses airtight pipelines to transport chemicals in water, which are then extracted by the receiver. Method 5 projects chemical odors into the air, where the receiver captures them using suction. Method 6 draws from optical fiber technology to transmit chemical odors through a cable, ensuring the chemical composition remains intact. Each method ensures the receiver experiences the same scent as the sender.

\end{abstract}


\IEEEpeerreviewmaketitle

\section{Introduction}

The ability to transmit sensory experiences, particularly smell, over long distances is an emerging frontier in communication technology. Unlike visual and auditory stimuli, which can be digitized and transmitted through existing mediums, the transmission of smell presents unique scientific and engineering challenges. Smells arise from volatile chemical compounds that interact with olfactory receptors in the nose, triggering distinct sensory perceptions. These odorants, or volatile compounds, are complex in composition and require novel approaches to effectively capture, encode, and reproduce at remote locations.

Recent advances in digital olfaction have opened new possibilities for encoding and transmitting smells. Research into electronic noses, for example, has demonstrated how molecular compositions of odors can be identified and categorized using machine learning models \cite{gutierrez2018electronic, li2019odorant}. These technologies lay the foundation for applications in virtual reality, healthcare, and communication systems. Similarly, wearable olfactory displays have enabled immersive experiences by synchronizing smell with other sensory inputs in augmented reality environments \cite{nagamatsu2019olfactory,karpeta2023umap}. Despite these advancements, the physical transportation and recreation of smells across distances remain underexplored.

This paper presents six innovative methods to address the challenge of smell transmission. These methods include the decomposition and reconstruction of chemicals using digital formulas, embedding chemicals into physical carriers such as water or ice, and leveraging pipelines or cables inspired by optical fiber technologies for direct transmission. Each approach focuses on preserving the chemical integrity and ensuring that the received smell closely matches the original. The proposed methods aim to bridge the gap between the digital encoding of smells and their tangible reproduction, drawing inspiration from advancements in both digital olfaction and molecular chemistry \cite{lindley2020digital,gutierrez2018electronic}.

By investigating these methods, this research contributes to the growing body of knowledge in digital olfaction and smell communication. It also identifies potential challenges, such as preserving chemical stability, preventing contamination during transport, and developing efficient reconstruction mechanisms. These findings offer a framework for further development of technologies that can expand the sensory dimension of communication and interaction across virtual and physical spaces.

\section{Methodology}

\subsection{Method 1: Decomposition and Reconstruction of Chemicals Using Chemical Formulas (Figure 1)}
In the first method, the chemicals will be decomposed, and we will determine their ingredients, such as their chemical formula. The information of this formula will then be sent through the internet to the receiver. On the receiver's side, to reconstruct the chemicals, the formula will be used. The ingredients of the chemicals, which already exist on the receiver's side, will be arranged in the same atomic structure and ratios as discovered on the sender's side. By using these same ingredients and ratios, the chemicals will be reconstructed on the receiver's side. Once this process is completed, the chemicals will form and present on the receiver's side.

\begin{figure} 
\centering
  \includegraphics[scale = 0.15]{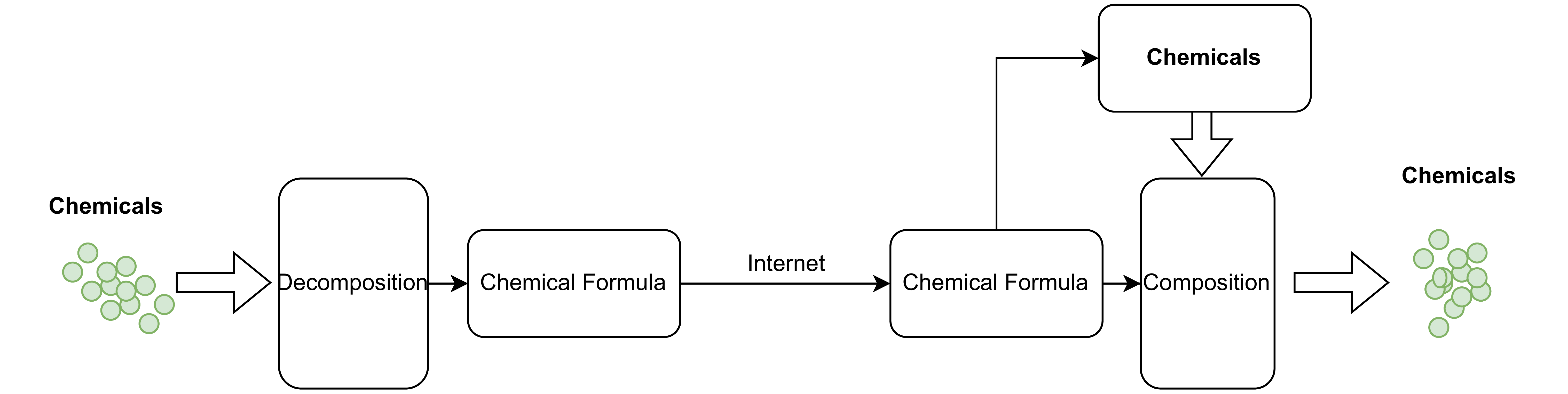}
  \caption{Decomposition and Reconstruction of Chemicals Using Chemical Formulas}
\label{f:Fig1}
\end{figure}

\subsection{Method 2: Chemical Embedding and Scent Transmission Through Ice Logistics (Figure 2)}

In this method, chemicals are embedded into water, which is then placed in an airtight container and frozen into ice. This process ensures that the molecules inside the water are also frozen and remain trapped in the ice until it melts or turns into steam. This happens on the sender's side using ice logistics. The ice or ice cubes are then transported from the sender to the receiver. Upon arrival, the receiver extracts the chemicals from the ice as it melts or vaporizes into steam (or other technological advanced methods). This steam contains the same chemicals that were embedded in the ice on the sender's side, allowing the receiver to smell it in the same way it was experienced by the sender. The molecules remain unchanged throughout the process, ensuring that the scent is identical.

\begin{figure}
\centering
  \includegraphics[scale = 0.13]{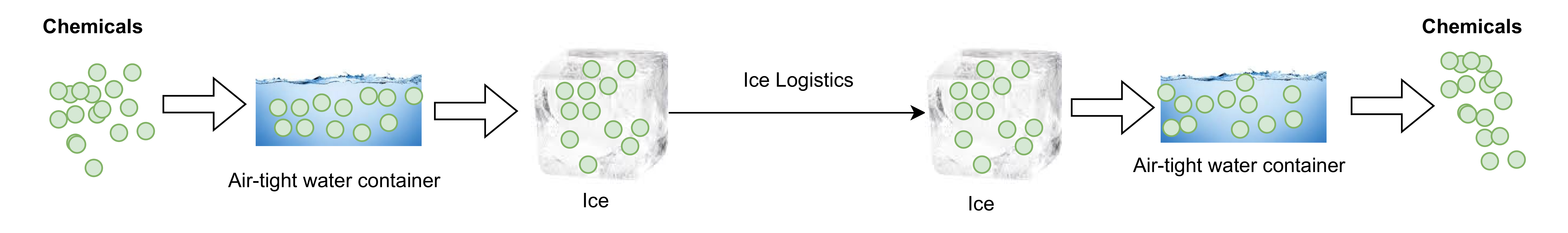}
  \caption{Chemical Embedding and Scent Transmission Through Ice Logistics}
\label{f:Fig3}
\end{figure}

\subsection{Method 3: Chemical Transfer and Scent Reproduction Through Water Logistics (Figure 3)}

In this method, chemicals are introduced into the water and then transported to the receiver. This process occurs on the sender's side using logistics. Upon arrival, the receiver extracts the chemicals from the water, which can be vaporized into steam (or other advanced technological methods). This steam contains the same chemicals that were embedded in the water on the sender's side, allowing the receiver to smell it in the same way as it was experienced by the sender. The molecules remain unchanged throughout the process, ensuring the scent is identical.

\begin{figure}
\centering
  \includegraphics[scale = 0.17]{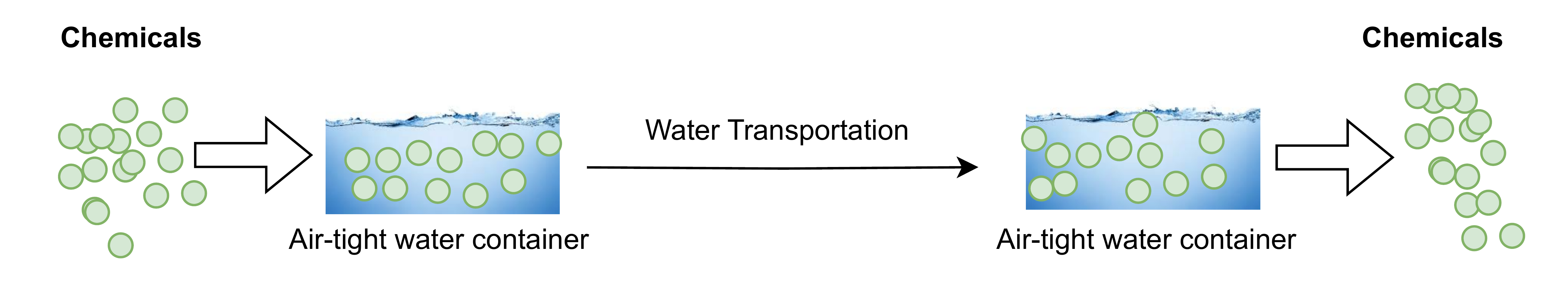}
  \caption{Chemical Transfer and Scent Reproduction Through Water Logistics}
\label{f:Fig2}
\end{figure}

\subsection{Method 4: Transport of Chemicals Through Water Pipelines for Smell Transmission (Figure 4)}

In this method, chemicals will be introduced into the water, and airtight pipelines will connect the sender and the receiver. Water will flow through this pipeline, carrying the chemicals. On the receiver's side, the chemicals will be extracted from the water, possibly by converting the water into steam or through other methods. Once extracted, the chemicals will be detectable, allowing the receiver to smell the same scent as the sender.

\begin{figure*}
\centering
  \includegraphics[scale = 0.17]{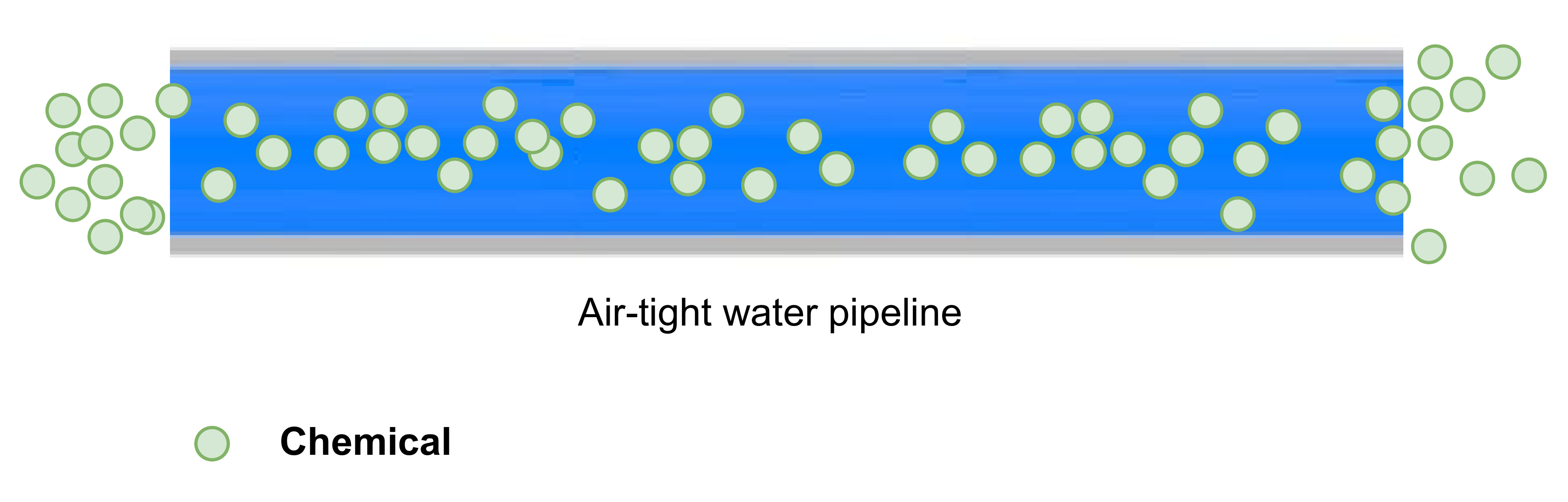}
  \caption{Transport of Chemicals Through Water Pipelines for Smell Transmission}
\label{f:Fig4}
\end{figure*}

\subsection{Method 5: Chemical Odor Projection and Capture Through Airborne Transfer (Figure 5)}

n this method, the chemical odors are projected into the air from the sender's side using a fan, blower, or other means. On the receiver's side, a suction or intake process will capture these chemical odors. This process may require maintaining a clear line of sight between the sender and receiver.

\begin{figure} 
\centering
  \includegraphics[scale = 0.17]{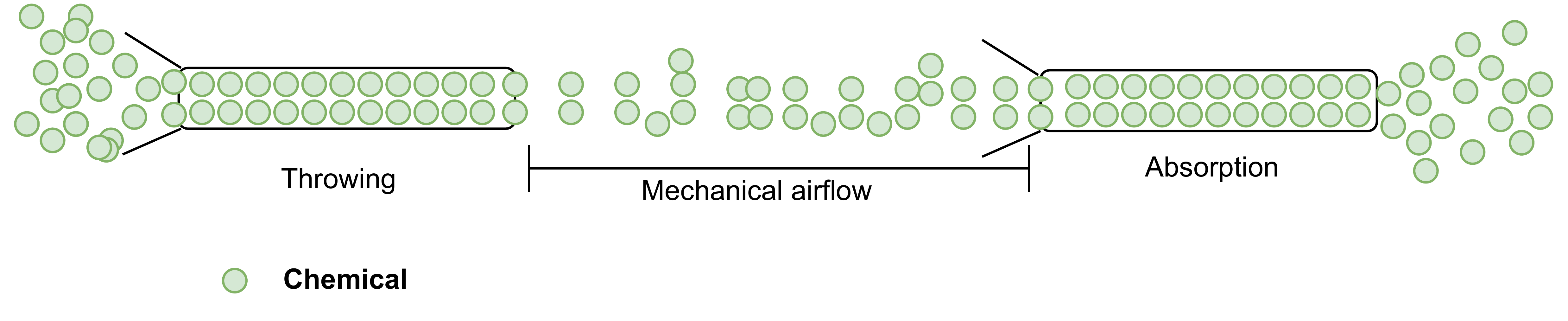}
  \caption{Chemical Odor Projection and Capture Through Airborne Transfer}
\label{f:Fig5}
\end{figure}

\subsection{Method 6: Transmitting Chemical Odors via Cable: A Fiber-Inspired Approach (Figure 6)}

The last method is more practical and is inspired by the concept of optical fibers, where light travels through a plastic or glass core. In this method, we hypothesize that chemical odors can be transmitted from the sender to the receiver using a similar approach. Instead of light, chemicals will travel through a core that supports the flow of air and the chemicals themselves, transmitted via a cable. The cable would function similarly to an optical fiber, guiding the chemical compounds from one location to another, ensuring that the chemical composition remains intact during the transfer. This method draws on existing technology to create a new approach for the transfer of odors.

\begin{figure}
\centering
  \includegraphics[scale = 0.17]{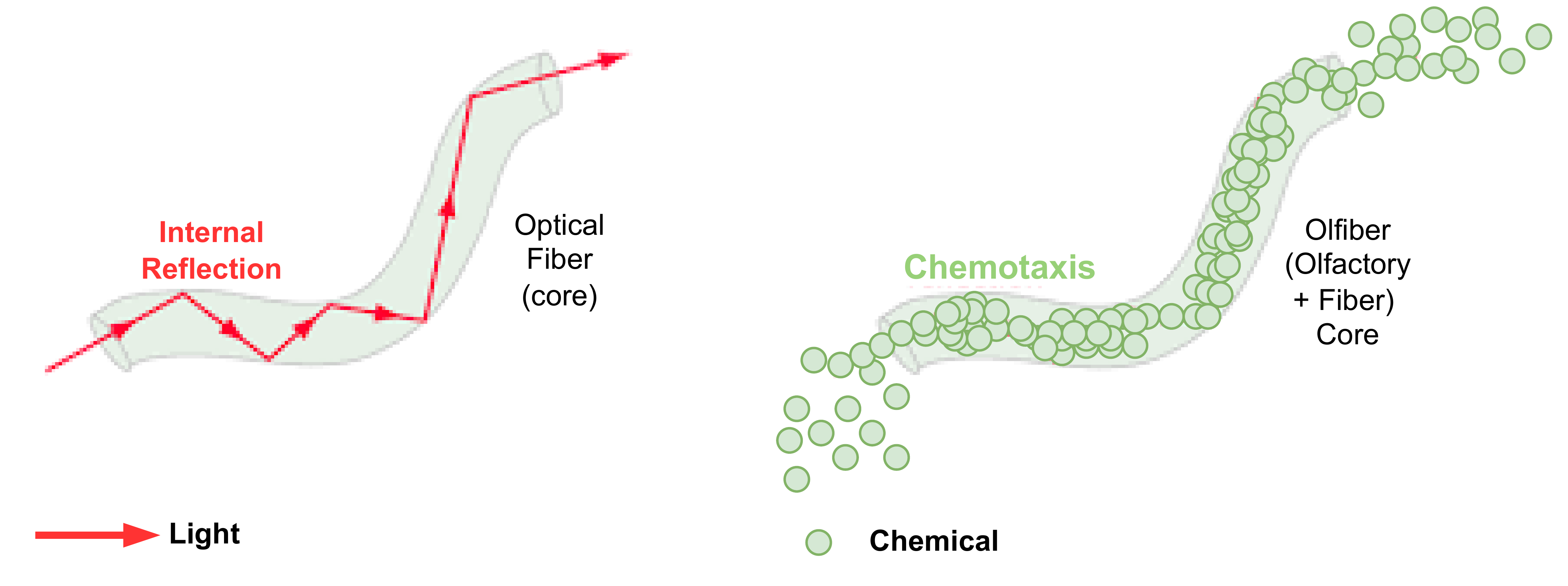}
  \caption{Transmitting Chemical Odors via Cable: A Fiber-Inspired Approach}
\label{f:Fig6}
\end{figure}

\section{Related Work}

Digital olfaction technologies have advanced significantly in recent years, aiming to analyze, transmit, and reproduce odors remotely. One prominent approach involves electronic noses that detect odorant molecules, analyze their composition, and digitally encode them for applications ranging from virtual reality to medical diagnostics \cite{gutierrez2018electronic}. These systems often use machine learning models to predict odorant perception from molecular features, providing insight into how smells are detected and encoded \cite{li2019odorant,karpeta2023umap}.

The integration of smell into virtual environments through olfactory augmented reality and wearable displays further highlights the practical applications of transmitting smells \cite{nagamatsu2019olfactory}. These devices aim to recreate odors in real time, drawing parallels to your method of reconstructing chemicals on the receiver side.

Furthermore, odorant classification research has demonstrated how the molecular structure and functional groups of odorants determine their perception. This aligns with your decomposition-reconstruction method, where the chemical formula is critical for transmitting the scent intact \cite{karpeta2023umap,li2019odorant}.

Finally, the concept of transmitting chemical odors through a physical medium, such as a cable or pipeline, echoes advancements in digital olfaction over networks. Studies have explored how odors can be encoded, transmitted, and decoded, similar to optical fibers used for data transmission \cite{lindley2020digital, gutierrez2018electronic}. These methods emphasize maintaining the integrity of the odorant molecules during transfer, aligning with your hypothesis of preserving chemical composition for accurate smell reproduction.

\section{Disclosure of AI Assistance}
This text has been assisted by a generative AI tool to enhance clarity. If you are already familiar with the background knowledge and related work, please focus on the methodology section, which is entirely my original contribution. The rest of the content has been structured to meet academic standards, supported and enhanced by generative AI tools to ensure completeness

\bibliographystyle{IEEEtran}

\bibliography{mybibfile}

\end{document}